\documentclass[preprint,12pt]{elsarticle}

\usepackage{amssymb}
\usepackage[utf8]{inputenc}
\usepackage[english]{babel}
\usepackage[usenames, dvipsnames]{color}

\journal{Nuclear Physics B}

\begin{document}

\begin{frontmatter}



\title{Average Energy Expended Per e-h Pair and Energy Scale Function
 for Germanium-Based Dark Matter Experiments}

\author[usd]{W.-Z. Wei}
\author[usd]{L. Wang}
\author[usd,yzu]{D.-M. Mei\corref{cor}}
\cortext[cor]{Corresponding author.}
\ead{Dongming.Mei@usd.edu}

\address[usd]{Department of Physics, The University of South Dakota, Vermillion, South Dakota 57069, USA}
\address[yzu]{School of Physics and Optoelectronic, Yangtze University, Jingzhou 434023, China}

\begin{abstract}
We report a new result of the temperature-dependent average energy expended per electron-hole (e-h) pair, $\varepsilon$, for germanium detectors. Applying energy partition mechanism in ionization for a given energy deposition, the Fano factor and the value of $\varepsilon$ can be determined separately. Subsequently, we illustrate the variation of $\varepsilon$ as a function of temperature. The impact of $\varepsilon$ on the energy threshold and energy scale for germanium detectors at a given temperature is evaluated. We demonstrate an absolute energy scale function of low-energy recoils for germanium detectors in the direct detection of dark matter particles.
\end{abstract}


\end{frontmatter}


\section{Introduction}
Many galactic observations indicate a large fraction ($\sim$85\%) of the total matter in the Universe is dark matter~\cite{dm1, dm2, dm3, dm4}. The most compelling candidate for dark matter particles is the WIMP (Weakly Interacting Massive Particle), which is believed to only interact through the weak force and gravity. Therefore, the interaction cross section with ordinary matter is extremely small. One of the ways to detect WIMPs is to measure the recoil energy of a nucleus from a WIMP-nucleus collision in a detector, either directly or indirectly, which will give an estimate of the mass of the incoming WIMPs~\cite{dan1}.

\begin{figure}
\centering
\includegraphics [angle=0,width=9.cm] {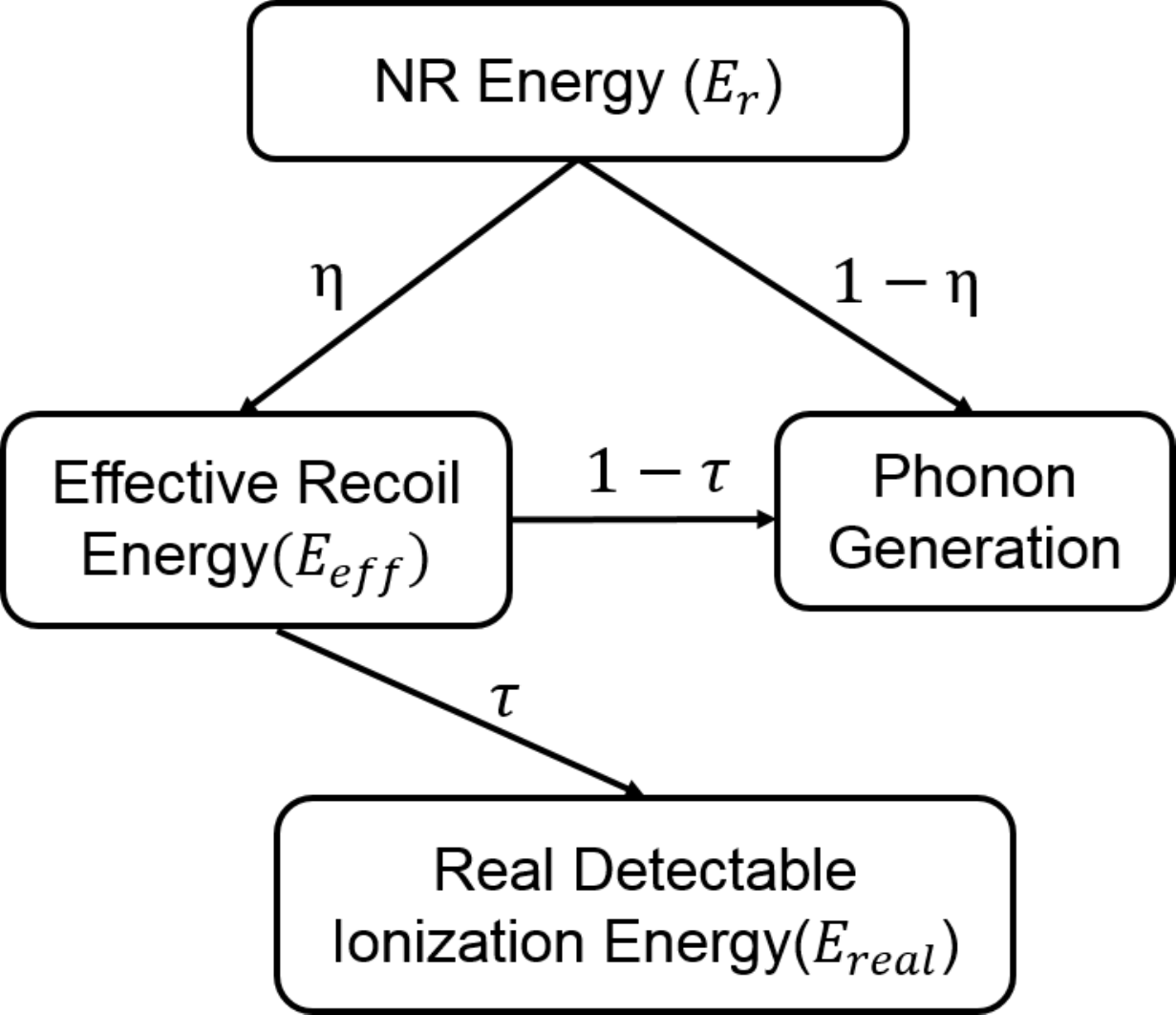}
\caption{\small{Energy deposition process in a Ge detector. Both $\eta$ and $\tau$ are ionization efficiency, which correspond to the energy loss process a) and b) mentioned in the introduction, respectively.}}
\label{fig:flow}
\end{figure}

The detector response to low-energy recoils plays a critical role in the direct detection of WIMPs. In a germanium (Ge) ionization detector, as shown in Fig.~\ref{fig:flow}, the energy loss from nuclear recoil energy ($E_r$) to the real detectable ionization energy ($E_{real}$) can be attributed to the following two main physics processes: a) the energy partition between ionization (electron-hole (e-h) pairs creation) and atomic motion (phonon generation) and b) the charge creation accompanying phonon generation due to momentum conservation in the creation of e-h pairs~\cite{mei}. The former process is described by the fraction of energy, $\eta$ in Fig.~\ref{fig:flow}, allocated to ionization. The latter process has assumed efficiency, $\tau$ in Fig.~\ref{fig:flow}, describing
energy loss to the creation of phonons in the creation of e-h pairs. Note that charge trapping was not taken into account in the process b) mentioned above since charge trapping is usually corrected by individual experiment using different ways before data are reported. Also note that charge recombination is not considered in the process b) as well since the energy reduction due to charge recombination is in a very low level for Ge detectors with sufficient applied field. Traditionally, only a) has been considered when developing an ionization efficiency model~\cite{lind,dan1,dan2} to compare with neutron calibration data in which the ionization efficiency is relative to the calibration from gamma-ray sources. This type of calibration is a relative energy calibration since the energy scale is determined using well-known gamma-ray energies on the assumption that the entire energy is detectable.

However, due to the additional energy loss process b) mentioned above, the real detectable ionization energy ($E_{real}$) is only a small fraction, $\frac{E_g}{\varepsilon}$ ($E_g$ is the band gap energy and $\varepsilon$ is the average energy required to produce an e-h pair), of the energy measured ($E_{vis}$) using the relative energy calibration for a given recoil event~\cite{mei}. 
This is because not all of the deposited energy is detectable as the creation of phonons that are inevitable in the creation of e-h pairs due to the required momentum conservation in process b)  and those phonons are not detectable by any generic Ge detectors. The energy loss to the creation of phonons in process b) can be estimated using the difference between $\varepsilon$ and $E_g$. Without the creation of phonons in process b), the expected number of e-h pairs can be expressed as: 
$N^{’}_{pairs}$ = $\frac{E_r\times\eta}{E_g}$. With the creation of phonons in process b), the detectable number of e-h pairs are $N_{pairs}$ = $\frac{E_r\times\eta}{\varepsilon}$. 
Taking the ratio of two formulas for the creation of number of e-h pairs yields, $N_{pairs}$ = $\frac{E_g}{\varepsilon}\times N^{’}_{pairs}$.  
Since the detected energy is proportional to $N_{pairs}$ and the deposited energy is proportional to $N^{’}_{pairs}$, the detectable energy can be expressed as $E_{real}$ = $\frac{E_g}{\varepsilon}\times E_r\times\eta$. 

In the relative energy scale, $E_{vis}$ = $E_r$$\times\eta$ = $\varepsilon$$\times$$N_{pairs}$, this is valid only if $\varepsilon$ is independent of energy for a given temperature. However, the adoption of $\varepsilon$ is to take into account the creation of phonons, which depends on the energy deposition processes, in the generation of e-h pairs. The energy deposition processes, for a given energy, can be photoelectric effect, Compton scattering, pair production, or the emission of Auger electrons, depending on the incoming particle type and its energy. Therefore, the creation of phonons has large fluctuation with respect to different energy deposition processes and $\varepsilon$ must be the average energy expended per e-h pair with tolerable uncertainty within a well-calibrated energy range. Beyond this range, in particular in the low energy region where the energy calibration cannot be implemented, the value of $\varepsilon$ may largely depend on the energy deposition processes~\cite{papp} and it cannot be easily understood.

Alternatively, $E_g$ is a constant and it is only related to the band structure of Ge for a given temperature. We can assume that the detectable energy is proportional to $N^{’}_{pairs}$ = $\frac{E_r}{E_g}$, where the reduction of e-h pairs due to the creation of phonons can be related to the energy deposition processes described by the ionization density, $dE/dx$,  which is energy and particle type dependent. 

For nuclear recoils, the calibration can be tricky and the measurements are widely spread in the low-energy region~\cite{dan1}. To interpret the difference among various measurements, one needs a model that takes into account both of the energy deposition processes a) and b) mentioned above. We developed such an absolute energy calibration model in our previous work~\cite{mei}. The real detectable ionization energy ($E_{real}$) in an absolute energy model for both electronic and nuclear recoils can be expressed as:
\begin{equation}
E_{real} = \frac{E_g}{\varepsilon}\times{E_{vis}} = \frac{\alpha}{1+\beta\frac{dE_{eff}}{dx}}\times \eta\times E_{r},
\label{eq:realEner}
\end{equation}
where $E_{vis}$ is the detectable energy in the relative energy scale, $\eta$ and $\frac{\alpha}{1+\beta\frac{dE_{eff}}{dx}}$ correspond to the energy reduction process a) and b) described above, respectively, $\eta$ can be calculated by the Lindhard theory~\cite{lind} or Barker-Mei model~\cite{dan1, dan2}, $\frac{\alpha}{1+\beta\frac{dE_{eff}}{dx}}$ was derived from Birks' law~\cite{birks1, birks2}, $\alpha$ and $\beta$ are both constants for a given temperature and can be determined experimentally, $\frac{dE_{eff}}{dx}$ is the stopping power that can be calculated for a given effective ionization energy, $E_{eff}=\eta\times E_r$, by using a theoretical model in our work~\cite{mei2}. In this work, an experiment utilizing Ge detector and known gamma-ray sources was designed to determine $\alpha$ and $\beta$ in the model.
Phonon generation exists in both process a) and b). Note that the double counting of phonon in the proposed model is avoided using the ionization efficiency, $\eta$ in the process a), and the e-h pair creation efficiency, $\tau$ = $\frac{\alpha}{1+\beta\frac{dE_{eff}}{dx}}$ in the process b), respectively. 

It is worth mentioning that Birks' law was originally proposed for estimating the light yield per path length as a function of the energy loss per path length for a particle traversing a scintillator. Since the light yield is only related to ionization density regardless of the type of detector, we can apply Birks' law to noble liquid detectors~\cite{mei2}. Note that the energy partition process in a Ge detector described in Fig.~\ref{fig:flow} is quite similar to that in a dual-phase liquid xenon detector, such as a LUX-like detector~\cite{lux}, for a given energy deposition. 

In a Ge detector or a LUX-like detector, Lindhard theory explains the energy loss from a nuclear recoil to ionization and atomic motion, which corresponds to the process a) mentioned above. In the ionization, the momentum conservation requires the yield of e-h pairs or e-ion pairs be accompanied by the generation of phonons~\cite{mei}. Since the light yield in a LUX-like detector is mainly from the recombination of e-ion pairs, Birks' law describes the effective energy of the e-ion production responsible for the photon production (S1 signal in a LUX-like detector) as a function of the ionization density, which corresponds to the process b) mentioned above.

Similarly, in a Ge detector, we can apply Birks' law for describing the effective energy of e-h pair production as a function ionization density, which also corresponds to the process b). The validation of the applicability of Birks' law to Ge detectors is verified in this work by comparing the absolute calibration model with all existing data.        

As indicated by Eq.~\ref{eq:realEner}, the variation of the average energy expended per e-h pair, $\varepsilon$, affects the variation of the real measurable energy ($E_{real}$) for a Ge detector. This necessitates a study on $\varepsilon$ especially its temperature dependence since two main types of Ge detectors are used in the direct detection of WIMPs: the generic Ge detectors with operating temperature around 77 Kelvin and the bolometer-type detectors with operating temperature at the milli-Kelvin range. $\varepsilon$ was accurately measured to be $\sim$3 eV at 77 Kelvin~\cite{E&R, shock,pehl, klein1, klein2, A&B}. However, there are no direct measurements for the value of $\varepsilon$ when the temperature decreases to the milli-Kelvin range. Therefore, measuring $\varepsilon$ at the milli-Kelvin range is needed for a bolometer-type detector since the value of $\varepsilon$ is critical to the determination of both energy scale (as can be seen in Eq.~\ref{eq:realEner}) and energy threshold for a bolometer-type detector~\cite{superCDMS}. It is quite difficult for a bolometer-type detector to measure the value of $\varepsilon$ without knowing the collection efficiencies of charge and three types of phonon (primary, recombination and Luke) in the energy deposition process in the detector~\cite{fallows}. Thus, an independent way to evaluate the value of $\varepsilon$ at the milli-Kelvin range is desired. As described in our previous work~\cite{mei}, $\varepsilon$ is related to the Fano factor through $F=\sqrt{\frac{E_x}{E_g}(\frac{\varepsilon}{E_g}-1)}$ with $E_x$ the average energy of primary phonon. The Fano factor ($F$) is also related to the detector energy resolution contributed by the intrinsic statistical variation ($\sigma_{stat}$) through $\sigma_{stat} =\sqrt{FN_i}\varepsilon =\sqrt{FE\varepsilon}$~\cite{fano} with $N_i = \frac{E}{\varepsilon}$. Thus, the value of $\varepsilon$ can be determined using the Fano factor ($F$) formula if the detector energy resolution due to the intrinsic statistical variation ($\sigma_{stat}$), the band gap energy ($E_g$) and the average primary phonon energy ($E_x$) are known.

In this paper, we present the study on $\varepsilon$ about its temperature dependence, its value at milli-Kelvin range and its impact on the energy threshold of a bolometer-type detector in section~\ref{s:aver}, followed by the investigation of energy scale calibration of Ge detectors in section~\ref{s:scale}. 
Finally, we summarize our conclusions in section~\ref{s:conc}.

\section{The average energy expended per e-h pair, $\varepsilon$}
\label{s:aver}
\begin{figure}
\centering
\includegraphics[angle=0,width=11.cm]{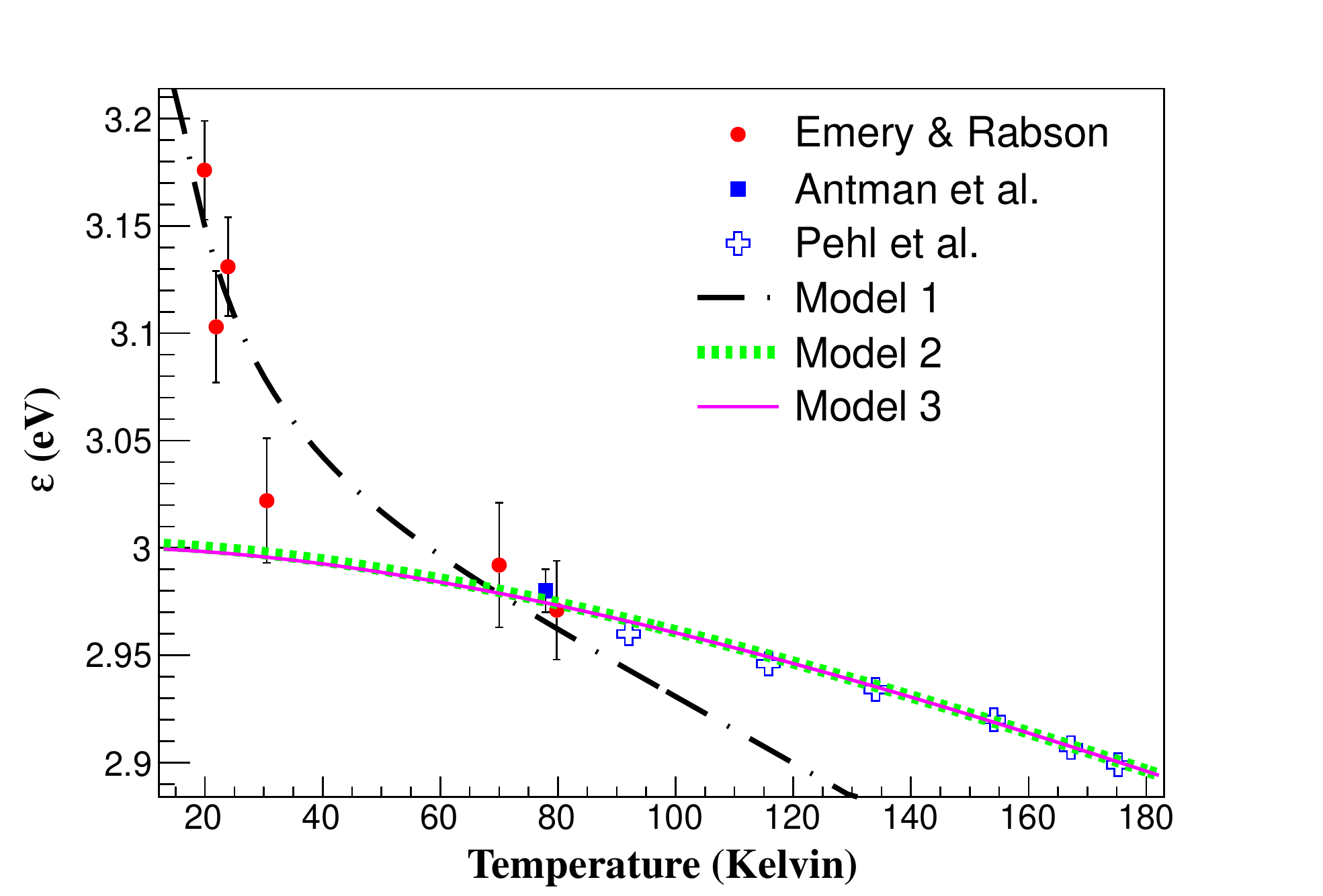}
\caption{\small{All existing data~\cite{E&R,pehl,antman} and theoretical models~\cite{E&R,klein1, klein2} for the variation of $\varepsilon$ with temperature. Note that the data points from~\cite{E&R} were taken after the corrections for charge trapping and recombination effects in the experiment~\cite{E&R}.}}
\label{fig:avgEner1}
\end{figure}

The temperature effects on $\varepsilon$ have been investigated for many years. Fig.~\ref{fig:avgEner1} shows the temperature dependence in $\varepsilon$ from existing data and theoretical models. The data were obtained by Emery and Rabson~\cite{E&R}, Pehl et al.~\cite{pehl}, and Antman et al.~\cite{antman}. To explain the data, based on the theory from Shockley~\cite{shock} ($\varepsilon = 2.2E_g+rE_R$), Emery and Rabson~\cite{E&R} developed a model (model 1 in Fig.~\ref{fig:avgEner1}): 
\begin{equation}
\varepsilon = 2.2\cdot E_g(T) + 1.99\cdot E_g(T)^{3/2}\cdot exp(\frac{4.75\cdot E_g(T)}{T}),
\label{avgE}
\end{equation}
where 1.99 and 4.75 are two fitting parameters, the band gap data, $E_g(T)$, are from Smith~\cite{smith}. Later on, 
also on the basis of Shockley theory~\cite{shock}, Klein~\cite{klein1, klein2} proposed another model:
\begin{equation}
\varepsilon = \frac{14}{5}E_g +r\hbar\omega_R,
\label{k_M}
\end{equation}
where $r\hbar\omega_R$ is the fraction of energy attributed to phonons, with $r$ the average number of phonons and $\omega_R$ the frequency of Raman phonon, which is the highest frequency among all optical phonons in the vibration. Model 2 and 3 in Fig.~\ref{fig:avgEner1} are the model in Eq.~\ref{k_M} with parameters provided by Varshni~\cite{varshni} and Thurmond~\cite{thurmond}, respectively. 

As can be seen in Fig.~\ref{fig:avgEner1}, none of the theoretical models are capable of explaining all existing data. To have a good interpretation of the variation of $\varepsilon$ as a function of temperature down to the milli-Kelvin range, it is necessary to evaluate the value of $\varepsilon$ at the milli-Kelvin range first and then find a model which can fit all data.  

\subsection{Estimation of $\varepsilon$ at 50 milli-Kelvin}
\label{s:esti}
For a bolometer-type detector, the operating temperature is $\sim$50 milli-Kelvin~\cite{cdmsprl,edelweiss}. The total measured phonon energy ($E_p$) of a bolometer-type detector is the sum of energies from three types of phonons~\cite{fallows,wang}: primary phonons, recombination phonons and Neganov-Trofimov-Luke phonons 
(often also called ``Luke phonons" for simplicity)~\cite{neg&tro, luke}. Primary phonons are produced due to displacements of nuclei and electrons. Recombination phonons are created at the electrodes due to the recombination of electrons and holes. The Luke phonons are generated due to the charge carriers drifting across the detector by the external applied electric field. Taking into account the detection efficiencies for charge carriers and each type of phonon, the total phonon energy ($E_p$) can be expressed as~\cite{ande}: 
\begin{equation}
E_p  = \eta_{pri}(E_r-\frac{E_Q}{\varepsilon}E_g)+\eta_{rec}(f_Q\frac{E_Q}{\varepsilon})E_g +\eta_{Luke}(f_Q\frac{E_Q}{\varepsilon}eV_b),
\label{ep}
\end{equation}
where $\eta_{pri}$, $\eta_{rec}$ and $\eta_{Luke}$ represent the detection efficiency for primary phonons, 
recombination phonons and Luke phonons, respectively, $f_Q$ is the fraction of the total charge observed, $E_r$ is the recoil energy, $E_Q$ is the ionization energy, $e$ is the elementary charge, and $V_b$ is the bias voltage. 

From Eq.~\ref{ep}, we can see that the mean energy expended per e-h pair, $\varepsilon$, cannot be determined if the four efficiencies $\eta_{pri}$, $\eta_{rec}$, $\eta_{Luke}$ and $f_Q$ are not known. Thus, an alternative way to estimate the value of $\varepsilon$ is needed.

Utilizing the energy partition process for an energy deposition in a Ge detector, we developed a theoretical model that relates $\varepsilon$ to the Fano factor ($F$), energy resolution ($\sigma_{stat}$) and average primary phonon energy ($E_x$) in our earlier work~\cite{mei}:
\begin{equation}
F=\sqrt{\frac{E_x}{E_g}(\frac{\varepsilon}{E_g}-1)},
\label{fano}
\end{equation}
and
\begin{equation}
\sigma_{stat} =\sqrt{FE\varepsilon},
\label{sig_stat}
\end{equation}
where the variation of $E_g$ with temperature can be evaluated by the following model~\cite{varshni, thurmond} based on the assumption that $E_g$ is proportional to $T$ at high temperatures and proportional to $T^2$ at low temperatures:
\begin{equation}
E_g (T) = 0.7437 - \frac{4.774\times{10^{-4}}\cdot T^2}{T + 235}
\label{Eg},
\end{equation}
where $E_g$ is in eV and $T$ is in Kelvin. 0.7437 is the value of $E_g$ at 0 Kelvin. This model is valid for all temperatures from 0 Kelvin to the melting point of Ge, $\sim$1211 Kelvin.

To solve Eq.~\ref{fano} and Eq.~\ref{sig_stat} for $\varepsilon$ and $F$, we need to investigate the average primary phonon energy, $E_x$, and the energy resolution due to the intrinsic statistical variation, $\sigma_{stat}$. 

\subsubsection{Determination of $E_x$}
The average primary phonon energy, $E_x$, has no temperature dependence since it mainly depends on the lattice type and spacing~\cite{E&R}. This indicates that we can use the value of $E_x$ at 77 Kelvin for the case of 50 milli-Kelvin.

At 77 Kelvin, $\varepsilon$ is almost a constant, $\sim$3 eV~\cite{E&R, shock, pehl, klein1, klein2, A&B}, 
$E_g$ = 0.73 eV from Eq.~\ref{Eg}, and $F$ = 0.13~\cite{klein1}. Substitute these values into Eq.~\ref{fano}, we can obtain $E_x$ = 0.00414 eV, which corresponds to $\sim$1 THz in terms of average frequency for the primary phonon. 

It is worth mentioning that the Raman phonon energy ($E_{R}$ = 0.037 eV) in Shockley's model~\cite{shock} 
($\varepsilon$ = 2.2$E_{g}$ + $rE_{R}$) and the average energy of primary phonon, $E_{x}$, 
determined using the measured Fano factor at 77 Kelvin, is different by a factor of $\sim$10. 
This is because $E_{R}$ and $E_{x}$ represent different type of phonons from the emission of primary phonons. Right after the primary phonons are generated by the recoiling particle, the primary phonons are very energetic and they down convert from the high-energy optical branch to the low-energy ($\sim$1THz) acoustic branch~\cite{fallows, wang}. Due to this decay process, it is the acoustic branch instead of optical branch of primary phonons that are the final state of phonons in the energy partition between ionization and lattice excitation, which determines the statistical variation (the Fano factor), for a given energy deposition. 
Furthermore, $E_{R}$ (Raman phonon energy) is the energy of optical phonons in the Raman vibration, 
which scatters the charge carriers capable of secondary ionizations during the thermalization process. 
While $E_{x}$ is the energy of the acoustic primary phonons in the final state of the energy partition between ionization and lattice excitation, i.e. $E_{0} = E_{i}N_{i} + E_{x}N_{x}$, where $E_{0}$ is the energy deposition of an incoming particle in the target, $E_{i}$ is the energy of e-h pairs per ionization, $N_{i}$ is the number of ionizations, and $N_{x}$ is the number of excitations. Note that each ionization leads to an e-h pair production accompanying with generation of phonons. 
Therefore, $N_{i}$ is the number of e-h pairs and $N_{x}$ is the number of phonons per ionization.  
Correspondingly, $E_{i}$ is the minimum energy (the indirect band gap energy) required for the production of 
a charge pair and $E_{x}$ is the average energy of phonons accompanying the production of an e-h pair.
Since the initial primary phonons, Raman phonons with energy $E_{R}$ (0.037 eV), are energetic optical phonons, they decay into acoustic phonons with an average energy of 0.00414 eV in the final state to participate in the energy partition between ionization and excitation for a given energy deposition, the values of $E_{R}$ and $E_{x}$ are different by a factor of $\sim$10.   

\subsubsection{Determination of $\sigma_{stat}$}
There are three main contributions to the total measured energy resolution, $\sigma_{tot}$, for a gamma-ray~\cite{HR}:
\begin{equation}
(\sigma_{tot})^2 = (\sigma_{noise})^2 + (\sigma_{stat})^2 + (\sigma_{in-ch})^2,
\label{sigma}
\end{equation}
where $\sigma_{noise}$, $\sigma_{stat}$ and $\sigma_{in-ch}$ are the energy resolution 
contributed by electronic noise, intrinsic statistical variation, and incomplete charge collection, respectively. 

SuperCDMS reported the measured Ge detector energy resolution in two papers~\cite{superCDMS,cdms_3}. 
One was reported in 2010~\cite{cdms_3}, which showed that the average detector energy resolution (ionization signal only) for energy below 10 keV is:
\begin{equation}
\sigma(E) = \sqrt{(0.293)^2 + (0.056)^2E},
\label{sig1}
\end{equation}
where the energy resolution, $\sigma(E)$, and the energy $E$ are both in keV. As discussed in our work~\cite{mei}, the term, $(0.056)^2E$, is the contribution mostly from the intrinsic statistical variation. According to Eq.~\ref{sigma}, we then have, $(\sigma_{stat})^2 = (0.056)^2E$. If one sets $\sigma_{stat}$ in this equation and Eq.~\ref{sig_stat} to be equal to each other, then we obtain, $\varepsilon F\times 10^{-3} = (0.056)^2$, where $\varepsilon$ is in eV and the factor of $10^{-3}$ is due to the unit conversion of $\varepsilon$ from eV to keV.

\begin{figure}
\centering
\includegraphics[angle=0,width=11.cm]{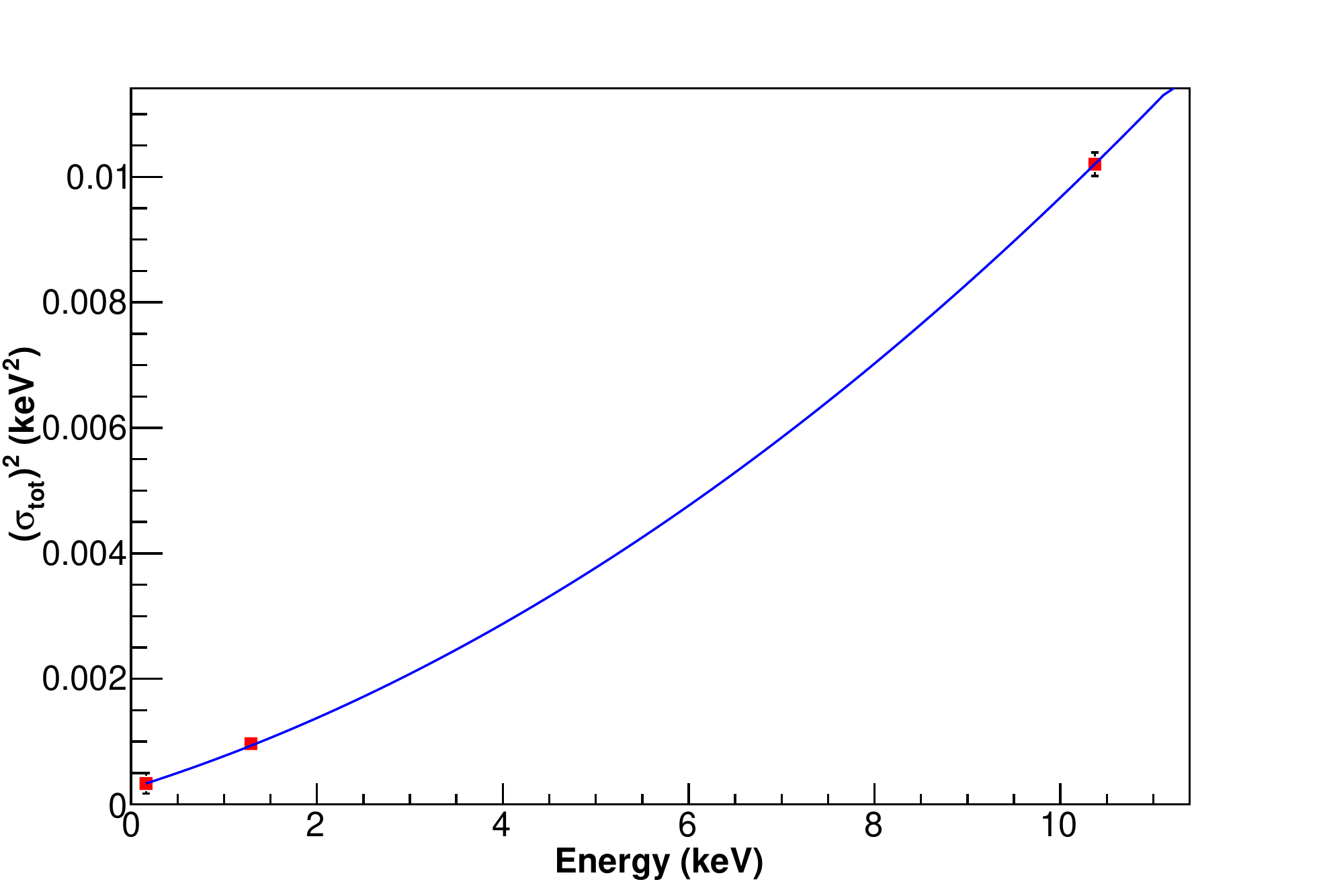}
\caption{\small{The energy resolution of CDMSlite detector~\cite{superCDMS} as a function of energy.}}
\label{fig:sigCDMS}
\end{figure}

The other energy resolution was reported in 2015~\cite{superCDMS}. According to Ref.~\cite{superCDMS}, the relative energy resolution ($\frac{\sigma}{\mu}$, $\mu$ is the peak energy) for three energy peaks from $^{71}$Ge electron-capture, 0.16 keV, 1.30 keV and 10.37 keV, are (11.4$\pm$2.8)\%, (2.36$\pm$0.15)\% and (0.974$\pm$0.009)\%, respectively. This allows us to generate $(\sigma_{tot})^{2}$ versus energy as presented in Fig.~\ref{fig:sigCDMS}. The best-fit function with reduced $\chi^2$ = 0 for data points in Fig.~\ref{fig:sigCDMS} is:
\begin{equation}
(\sigma_{tot})^2  = (2.57\pm1.90)\times10^{-4}+ (4.64\pm2.04)\times10^{-4}E+ (4.77\pm1.83)\times10^{-5}E^2, 
\label{sigCDMS2}
\end{equation}
where, $E$ is the energy in keV. The constant, linear and quadratic terms in Eq.~\ref{sigCDMS2} correspond
 to $(\sigma_{noise})^2$, $(\sigma_{stat})^2$ and $(\sigma_{in-ch})^2$ in Eq.~\ref{sigma}, respectively. Hence, we have,
$(\sigma_{stat})^2 = (4.64\pm2.04)\times10^{-4}E$, where E is the energy in keV. Similarly, we set $\sigma_{stat}$ in this equation and Eq.~\ref{sig_stat} to be equal to each other and we obtain, $\varepsilon F = 0.464\pm0.204$, where $\varepsilon$ is in eV.

\subsubsection{Determination of $\varepsilon$}
Using Eq.~\ref{fano} and substituting $E_g$ and $E_x$ with their value, 0.74 eV (Eq.~\ref{Eg}) and 0.00414 eV, respectively, we obtain, $F=\sqrt{5.59\times10^{-3}\cdot(\frac{\varepsilon}{0.74}-1)}$, where $\varepsilon$ is in eV. If one combines this equation with $\varepsilon F\times 10^{-3} = (0.056)^2$ and solves for $F$ and $\varepsilon$, we have $F$ = 0.28 and $\varepsilon$ = 11.3 eV. However, if one combines this equation and $\varepsilon F = 0.464\pm0.204$ together, we have $F$ = 0.14$^{+0.02}_{-0.03}$ and $\varepsilon$ = 3.32$^{+0.84}_{-0.97}$ eV. Two different sets of values indicate that the determination of $\varepsilon$ and Fano factor ($F$) using this method depends strongly on the measured energy resolution in which the noise contribution must be largely separated from statistical variation. Note that the value of $\varepsilon$ = 11.3 eV is not reasonable since the statistical variation is probably overestimated by assuming the linear term in the measured energy resolution shown in Eq.~\ref{sig1} is totally contributed by the statistical variation~\cite{lauren}. Other broadening effects such as time variance, position variance, charge collection on the detector resolution could also contribute to the linear term in Eq.~\ref{sig1}~\cite{lauren}. The value of $\varepsilon$ = 3.32$^{+0.84}_{-0.97}$ eV is more reasonable
since the energy resolution was optimized~\cite{lauren}. 

\subsubsection{Comparison with measured $\varepsilon$ at 50 milli-Kelvin}
\begin{figure}
\centering
\includegraphics[angle=0,width=9.5cm]{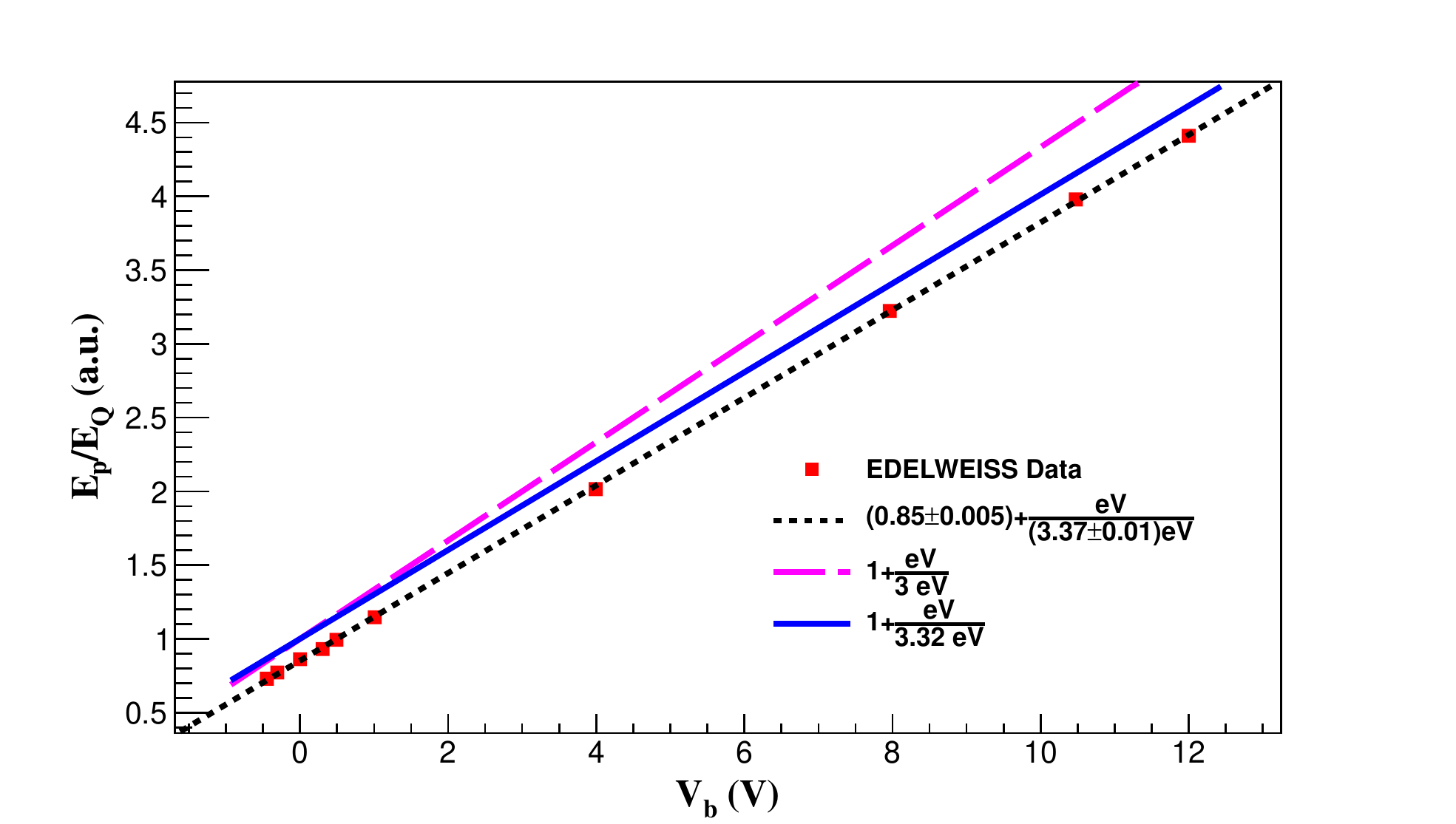}
\caption{\small{The ratio of total phonon energy ($E_p$) to ionization energy ($E_Q$) versus bias voltage ($V_b)$. The data (red points) were measured by the EDELWEISS detector~\cite{edelweiss2}.}}
\label{fig:g_v}
\end{figure}

The value of $\varepsilon$ at 50 milli-Kelvin was measured by the EDELWEISS dark matter experiment~\cite{edelweiss2} using the relationship, $E_p = E_Q(1+\frac{eV_b}{\varepsilon})$ (for electronic recoils), since the ratio of total phonon energy ($E_p$) to ionization energy ($E_Q$) can be measured for a given bias voltage ($V_b$) as shown in Fig.~\ref{fig:g_v}. The detailed work about how to obtain the above relationship ($E_p = E_Q(1+\frac{eV_b}{\varepsilon})$) for electronic recoils from Eq.~\ref{ep} will be discussed in section~\ref{s:thres}. 
 
As shown in Fig.~\ref{fig:g_v}, $\varepsilon$ = 3.32 eV (blue solid line) has a better agreement with data (red points) than $\varepsilon$ = 3.0 eV (magenta dashed line). The best fit (black dashed line) indicates that $\varepsilon$ = (3.37$\pm$0.01) eV, which also verifies that $\varepsilon$ = 3.32 eV measured using the energy resolution from SuperCDMS at 50 milli-Kelvin. The physics meaning of 0.85 in the best fit is the conversion efficiency from primary and recombination phonons to thermal phonons relative to the conversion efficiency from Luke phonons to thermal phonons. 

\subsection{The variation of $\varepsilon$ with temperature}
\label{s:tempD}
To obtain a reasonable value of $\varepsilon$ at 50 milli-Kelvin, it is necessary to develop a model of $\varepsilon$ as a function of temperature. This model can fit all existing data shown in Fig.~\ref{fig:avgEner1} and also can predict the value of $\varepsilon$ at 50 milli-Kelvin. Using the theory from Emery and Rabson~\cite{E&R} presented in Eq.~\ref{avgE}, according to the definition of $\varepsilon$, the first term in Eq.~\ref{avgE} consists of two components: a) the band gap energy, $E_g$, and b) the final retained kinetic energy ($E_f$) of electrons and holes which cannot further conduct ionization in the detector with assumption that $\sim$60\% of $E_g$ is retained by both electrons and holes, i.e., $E_f\sim$ 0.6$E_g$. Hence, Eq.~\ref{avgE} can be rewritten as:
\begin{equation}
\varepsilon = E_g(T) + 2\cdot0.6E_g(T)+ B\cdot E_g(T)^{3/2}\cdot exp(\frac{C\cdot E_g(T)}{T}),
\label{step1}
\end{equation}
where $B$ and $C$ are constants and needed to be determined from data. There are two issues in this model when fitting all data as shown in Fig.~\ref{fig:avgEner1}: a) it cannot explain the data in the high temperature range, $T >$ 80 Kelvin; b) it will blow up when $T$ is close to zero. The first issue is because the second term in Eq.~\ref{step1}, the final retained kinetic energy ($E_f$) of charge carriers is dependent on the Auger recombination-impact ionization process which has strong temperature dependence~\cite{E&R, B&L}. However, only assuming $E_f\sim$ 0.6$E_g$ cannot provide sufficient temperature effects on $E_f$ to explain data in the high temperature range since the temperature dependence in $E_g(T)$ is too weak as can be seen from Eq.~\ref{Eg}. Thus, an additional temperature factor needs to be added into the second term in Eq.~\ref{step1}. The second issue is due to no constraint in the denominator of the last term in Eq.~\ref{step1}, so that $\varepsilon$ will reach infinity when $T$ is approaching zero. 
Therefore, to resolve these two issues, we made two corrections in Eq.~\ref{step1} and then it becomes: 
\begin{equation}
\varepsilon = E_g(T) +1.2\cdot E_g(T)\cdot T^{A} +B\cdot E_g(T)^{3/2}\cdot exp(\frac{C\cdot E_g(T)}{T+D}), 
\label{avgE2}
\end{equation}
where both $B$ and $C$ are constants and needed to be determined from data. $A$ and $D$ are chosen values that give the best fit for all data points except points at 3.32$^{+0.84}_{-0.97}$ eV and 11.3 eV for 50 milli-Kelvin.

\begin{figure}
\centering
\includegraphics[angle=0,width=11.cm]{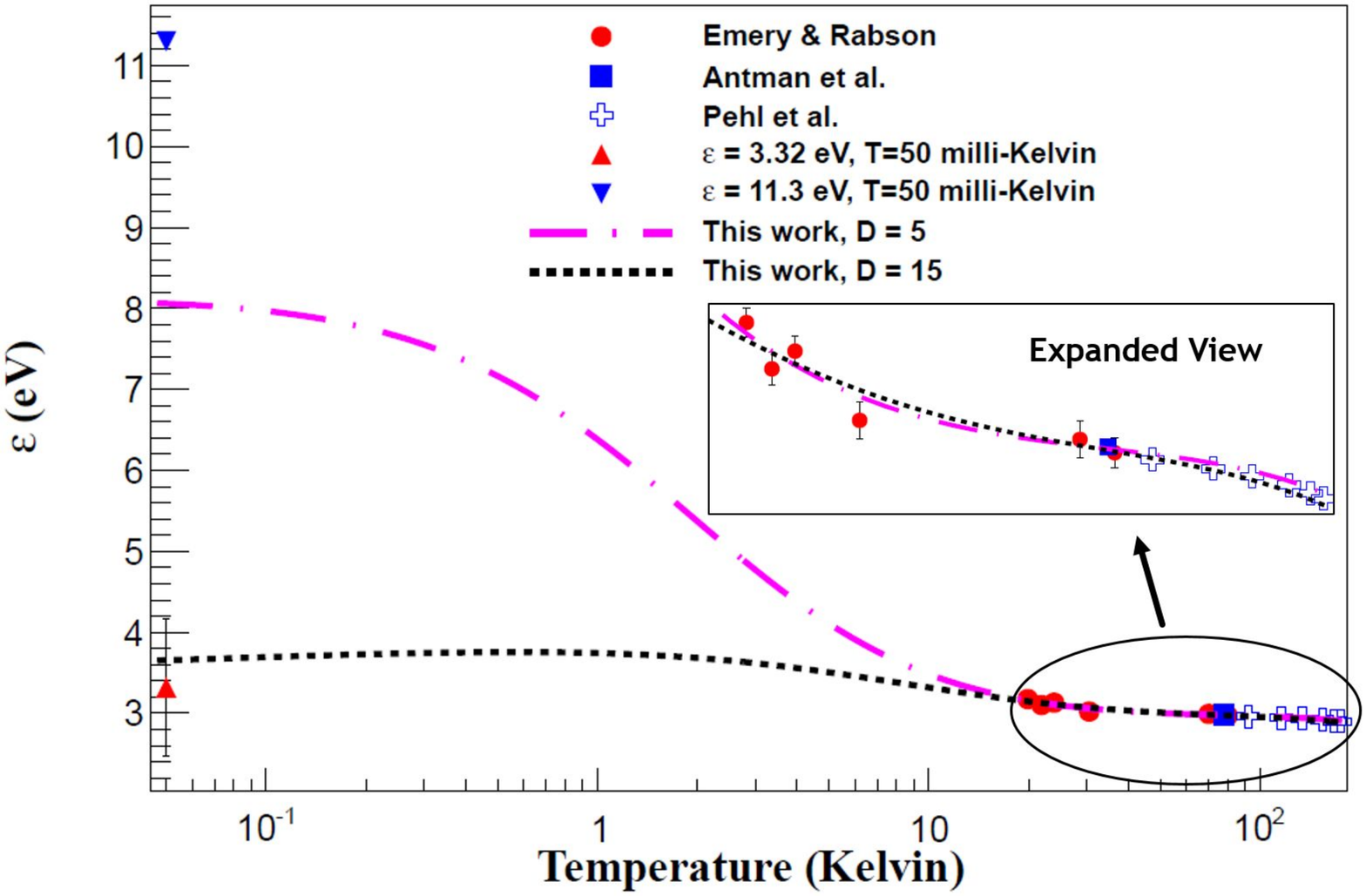}
\caption{\small{$\varepsilon$ versus temperature with all available data points~\cite{E&R,pehl,antman} and calculated value at $T$ = 50 milli-Kelvin.}}
\label{fig:avgEner2}
\end{figure}

As shown in Fig.~\ref{fig:avgEner2}, the model in Eq.~\ref{avgE2} fits well all existing data with $E_g$ 
in the form of Eq.~\ref{Eg}, $A$ = 0.1 and $D$ = 5 $\sim$ 15. $B$ = 1.23, $C$ = 14.48 for $D$ = 5, and $B$ = 1.17, $C$ = 22.22 for $D$ = 15. Due to the uncertainty of $D$, the model in Eq.~\ref{avgE2} can only predict the range of the value of $\varepsilon$, which is 3.55 $\sim$ 7.95 eV. This range verifies that there is overestimate in the value of $\varepsilon$ = 11.3 eV and the value of $\varepsilon$ = 3.32$^{+0.84}_{-0.97}$ eV is close to the lower value of the allowed range. To more precisely predict the value of $\varepsilon$ at 50 milli-Kelvin by using the model in Eq.~\ref{avgE2}, more measurements of $\varepsilon$ for low temperatures (T$<$20 Kelvin) are needed to decrease the uncertainty in $D$. 

\subsection{The impact of $\varepsilon$ on the energy threshold of a bolometer-type detector}
\label{s:thres}
The nuclear recoil energy threshold, $E_r$, for experiments using bolometer-type detectors is closely related to $\varepsilon$ according to Eq.~\ref{ep}. However, Eq.~\ref{ep} cannot be directly applied to a bolometer-type detector, which can be only sensitive to thermal phonons. Under the assumption that all phonons, converted into thermal phonons, are being detected at 100$\%$ efficiency by bolometer-type detectors~\cite{fallows}, Eq.~\ref{ep} reduces to be the ideal case~\cite{superCDMS,fallows}:
\begin{equation}
 E_p = E_r + \frac{eV_b}{\varepsilon}E_Q,
\label{eq14a}
\end{equation}
where the ionization energy ($E_Q$) is related to the recoil energy ($E_r$) through $E_Q \equiv E_r\cdot \eta$ ~\cite{fallows} 
with $\eta$ the ionization yield or the so-called ionization efficiency. Hence, Eq.~\ref{eq14a} can be rewritten as:
\begin{equation}
E_p = E_r(1 + \frac{eV_b}{\varepsilon}\eta),
\label{eq14_1}
\end{equation}
where $\eta \equiv$ 1 for electronic recoils. For nuclear recoils, $\eta$ can be calculated by the Lindhard theory~\cite{lind} or Barker-Mei model\cite{dan1, dan2}. With $V_b$ = 69 Volts and $\varepsilon$ = 3.0 eV, the nuclear recoil equivalent energy threshold ($E_r$) reported by SuperCDMS is~\cite{cdms_2},
$E_r = 2$ keV, which corresponds to $E_p$ = 10.7 keV according to Eq.~\ref{eq14_1}. 

Nevertheless, with the total phonon energy ($E_p$) measured to be 10.7 keV and $V_b$ = 69 Volts, 
the energy threshold of a bolometer-type detector will be about 7.5\% different if the value of $\varepsilon$ varies from 3.0 eV to 3.32 eV. This is to say, the value of $\varepsilon$ at 50 milli-Kelvin is critical to the determination of nuclear recoil threshold for a bolometer-type experiment. Therefore, a direct measurement of $\varepsilon$ needs to be performed at the level of 50 milli-Kelvin.

\section{Ge detector energy scale calibration}
\label{s:scale}
  
\subsection{Ge detector response to low-energy recoils}
 With consideration of the two main energy loss processes described in the introduction, we developed an absolute energy calibration model (Eq.~\ref{eq:realEner}), which has been described in detail in our previous work~\cite{mei}. As indicated by Eq.~\ref{eq:realEner}, whether the detector response to electronic recoils is linear or non-linear depends on if the term, $\beta\frac{dE_{eff}}{dx}$, can be ignorable or not. Since for low-energy electronic recoils, $\eta\equiv$ 1 and if $\beta\frac{dE_{eff}}{dx}$ is so small that it can be neglected. Eq.~\ref{eq:realEner} becomes:
\begin{equation}
E_{real} = \frac{E_g}{\varepsilon}\times{E_{vis}} = \alpha\times E_{r},
\label{eq17}
\end{equation}
where $E_g$, $\varepsilon$ and $\alpha$ are all constants and hence can be normalized to be a unit. 
In this case, the detector response to low-energy electronic recoils is linear. Otherwise, $E_{real}$ is a non-linear function of $E_r$ even for electronic recoils.

For low-energy nuclear recoils, $\eta$ is a function of recoil energy ($E_r$), and $\beta\frac{dE_{eff}}{dx}$ cannot be ignored due to the significant value of $\frac{dE_{eff}}{dx}$. Therefore, the detector response to low-energy nuclear recoils should be non-linear, which is in the form of Eq.~\ref{eq:realEner} with absolute ionization efficiency:
\begin{equation}
\eta_{tot}=\frac{\alpha}{1+\beta\frac{dE_{eff}}{dx}}\times \eta. 
\label{eq16}
\end{equation}

\subsubsection{Determination of $\alpha$ and $\beta$ at T = 77 Kelvin}
\begin{figure}
\centering
\includegraphics [angle=0,width=11.cm] {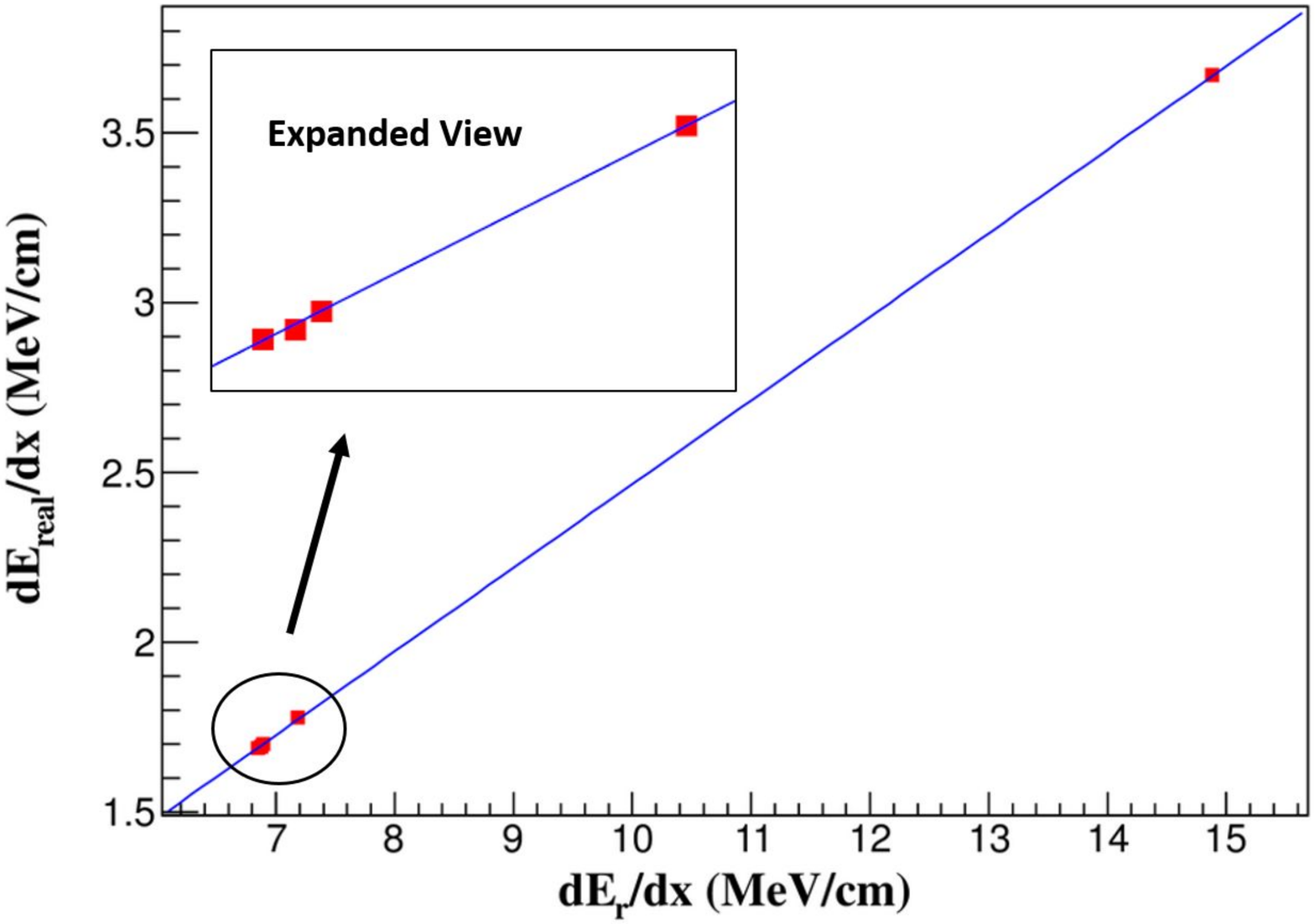}
\caption{\small{Determination of $\alpha$ and $\beta$ through known gamma rays.}}
\label{fig:ab}
\end{figure}

The energy reduction term, $\frac{\alpha}{1+\beta\frac{dE_{eff}}{dx}}$, in Eq.~\ref{eq:realEner} was derived from Birks' Law~\cite{birks1, birks2}:
\begin{equation}
\frac{dE_{real}}{dx} =  \frac{\alpha}{1+\beta\frac{dE_{eff}}{dx}}\frac{dE_{eff}}{dx},
\label{eq18}
\end{equation}
where $E_{real}$ = $\frac{E_g}{\varepsilon}\times{E_{vis}}$ from Eq.~\ref{eq:realEner}. For ERs, $E_{eff} = E_{r}$ since $\eta\equiv$ 1. Thus, in the case of electronic recoils, Eq.~\ref{eq18} becomes:
\begin{equation}
\frac{dE_{real}}{dx} =\frac{E_g}{\varepsilon}\frac{dE_{vis}}{dx} =  \frac{\alpha}{1+\beta\frac{dE_{r}}{dx}}\frac{dE_{r}}{dx},
\label{eq19}
\end{equation}
which corresponds to Eq. 3 in our work~\cite{mei}. From Eq.~\ref{eq19}, we can see that $\alpha$ and $\beta$ can be measured through electronic recoils. Since for a given known gamma ray, $E_{r}$ is the known recoil energy. $\frac{dE_{r}}{dx}$ can then be obtained through the database~\cite{nist} with known $E_{r}$.
To obtain $\frac{dE_{real}}{dx}$, one can measure $dE_{vis}$ and ${dx}$ in Eq.~\ref{eq19}. Once $dE_{vis}$ and $dx$ are known, $E_g$ = 0.73 eV and $\varepsilon$ = 3 eV for $T$ = 77 Kelvin , then $\frac{dE_{real}}{dx}$ can be calculated through the first part of Eq.~\ref{eq19}. With this idea, $^{109}$Cd, $^{22}$Na and $^{60}$Co were chosen to generate five gamma rays with energies: 88 keV ($^{109}$Cd), 511 keV and 1275 keV ($^{22}$Na), 1173 keV and 1333 keV ($^{60}$Co). To detect these gamma rays, a coaxial Ge detector from Princeton Gamma Tech with model RG11B/C~\cite{prin} was used in our experiment. A National Instruments PXI-1031 system~\cite{pxi} and Igor Pro 4.07 software~\cite{igor} were used for our data acquisition. With this experimental setup, $dE_{vis}$ for each gamma ray was measured using pulse shape analysis where the multiple site events and single site events can be distinguished in a similar way to Majorana~\cite{maj} and GERDA~\cite{gerda}. With $dE_{vis}$ measured from pulse shape analysis, the corresponding path length $\delta x$ can be derived through the database~\cite{nist}. The average path length, $dx$, for a given gamma ray energy, is determined through $dx$=$\frac{\sum{n_{i}\delta x}}{\sum{n_{i}}}$, where $n_{i}$ is the $i$th pulse in an event. Since $\frac{dE_{real}}{dx}$ is known,  
$\alpha$ and $\beta$ can be determined by fitting those five data points in the plot of $\frac{dE_{real}}{dx}$ as a function of $\frac{dE_{r}}{dx}$, as shown in Fig.~\ref{fig:ab}.
The best-fit function with reduced $\chi^2$ = 21.5 for the data points in Fig.~\ref{fig:ab} is:
\begin{equation}
\frac{dE_{real}}{dx} =  \frac{0.247\pm2.24\times{10^{-4}}}{1+(5.66\times{10^{-5}}\pm4.55\times{10^{-6}})\frac{dE_{r}}{dx}}\frac{dE_{r}}{dx}.
\label{eq20}
\end{equation}
By comparing Eq.~\ref{eq20} with Eq.~\ref{eq19}, we have, $\alpha = 0.247\pm2.24\times{10^{-4}}$ and
$\beta = 5.66\times{10^{-5}}\pm4.55\times{10^{-6}}$, which agree with our theoretical prediction~\cite{mei}, $\alpha$ = (0.249 $\pm$ 0.013) 
and $\beta$ = (5.12 $\pm$ 2.68) $\times$ 10$^{-5}$, within a reasonable range. Note that $\alpha$ has temperature dependence since $\alpha$ = $\frac{E_g}{\varepsilon}$ from Eq.~\ref{eq17} and both $E_g$ and and $\varepsilon$ have temperature dependence as discussed in section~\ref{s:aver}. However, $\beta$ has no temperature dependence because $\beta$ is only related to the stopping power $\frac{dE}{dx}$, which has nothing to do with temperature. 

\subsection{Absolute energy calibration vs. relative energy calibration}
\begin{figure}
\centering
\includegraphics[angle=0,width=11cm]{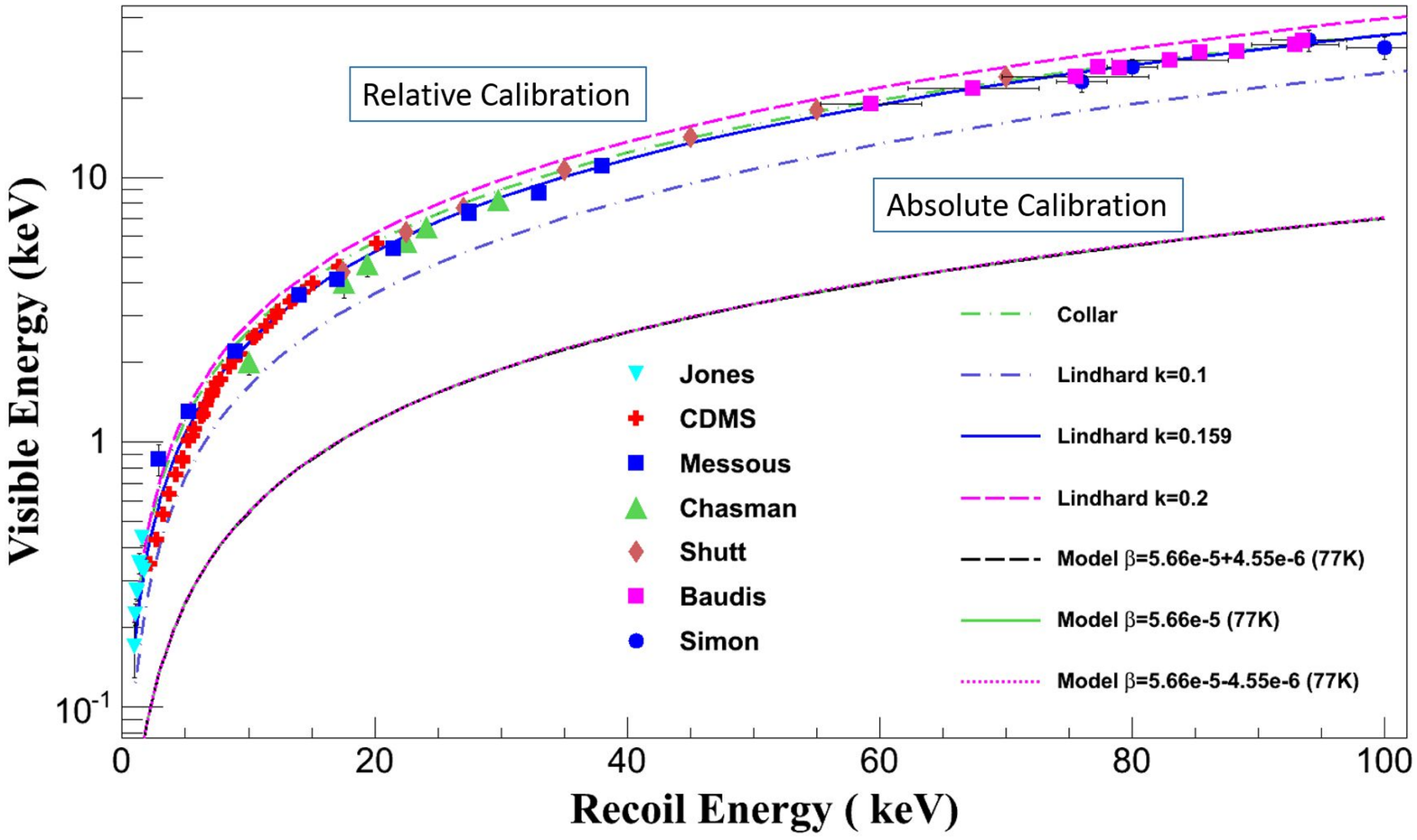}
\caption{\small{Comparison between absolute and relative energy calibration~\cite{lind,cdms_2,jones,collar,mess,chas,shutt,baud,simon}.}}
\label{fig:com1}
\end{figure}

The comparison between our absolute energy calibration model and all other existing data 
and models with a relative energy calibration at $T$ = 77 Kelvin is shown in Fig.~\ref{fig:com1}. Fig.~\ref{fig:com1} indicates that, for the same recoil energy, the eventual detectable ionization energy using absolute calibration is smaller than that of using a relative calibration. This means that the absolute ionization efficiency is smaller than the relative ionization efficiency for a given recoil energy.  
Note that the uncertainty of $\beta$ was taken into account in our absolute calibration model in Fig.~\ref{fig:com1}. 
In addition to the difference in ionization efficiency, another important difference is that the absolute energy calibration possesses the prediction power for the entire energy region beyond the data points covered the region from energy source calibration, while the relative energy calibration is only valid for the region that is fully calibrated using energy sources. Any extrapolation beyond the region of the calibration will result in uncertainty in the energy scale due to the lack of considering all energy deposition processes
in any data-driven model. This difference is because the absolute energy calibration has considered all possible energy deposition processes for a given electronic or nuclear recoil event in the detector. Therefore, the absolute energy calibration is more accurate and reliable than the relative energy calibration.

\subsection{Comparison between the existing data and the absolute calibration model}
\begin{figure}
\centering
\includegraphics[angle=0,width=11.cm]{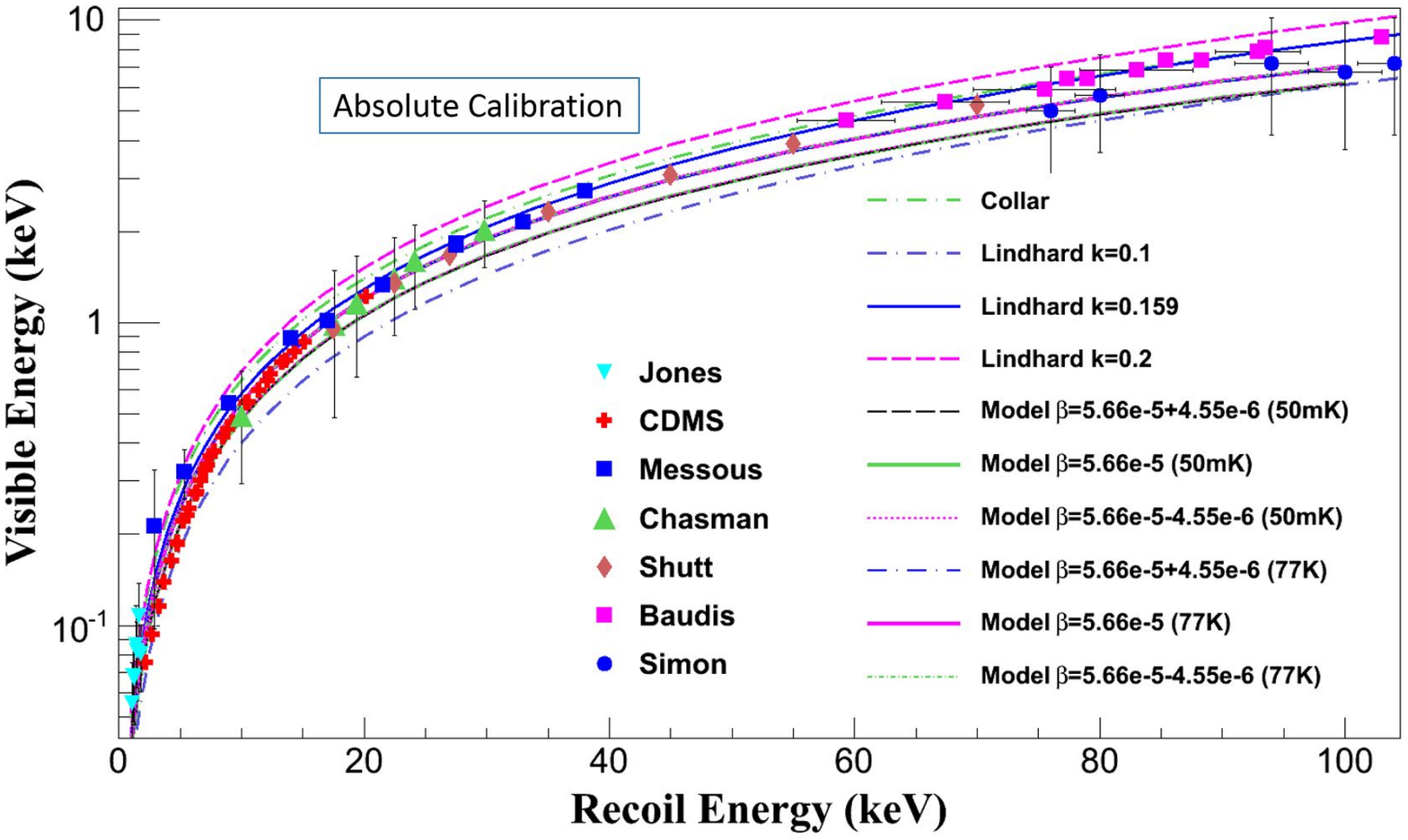}
\caption{\small{Comparison among all existing data and models with absolute energy calibration~\cite{lind,cdms_2,jones,collar,mess,chas,shutt,baud,simon}.}}
\label{fig:com2}
\end{figure}

One can apply the absolute calibration model to all existing experimental data and theoretical models to see if there is an agreement between them for recoil energy from 1 to 100 keV. Since all existing data and models adopt the relative calibration, we multiplied their measured visible energy by an additional factor, $\frac{E_{g}}{\varepsilon}$ ($\varepsilon$ = 3.0 eV for 77 Kelvin and $\varepsilon$ = 3.32 eV for 50 milli-Kelvin),  which takes into account the missing energy through the emission of phonons, to apply the absolute calibration model and get the real detectable ionization energy. As shown in Fig.~\ref{fig:com2}, our model agrees well with other data and models at $T$ = 50 milli-Kelvin for SuperCDMS~\cite{cdms_2}, Shutt~\cite{shutt} and Simon~\cite{simon}, and $T$ = 77 Kelvin for the rest. This verifies that Birks' law can be applied to Ge detectors.

\section{Conclusion}
\label{s:conc}
Traditionally, the product of $\varepsilon F$ is measured using the energy resolution function for a given detector. To determine the value of Fano factor or the value of $\varepsilon$, one has to assume either Fano factor or $\varepsilon$ is a constant. In this paper, we develop a method that allows the average energy expended per e-h pair, $\varepsilon$, and the Fano factor to be measured separately. Using the measured
energy resolution functions reported by SuperCDMS experiment, the values of $\varepsilon$ and Fano factor are determined for a detector operated at 50 milli-Kelvin. We demonstrate that our method depends strongly on the energy resolution function in which the statistical term is well defined. Using the existing data, we illustrate the best fit function that predicts the range of $\varepsilon$, which is within the range from the SuperCDMS measurements at 50 milli-Kelvin. 
According to the best fit of data and the best agreement with the model prediction, the value of $\varepsilon$ is 3.32 eV.
The energy threshold of a bolometer-type detector will be about 7.5\% different if the value of $\varepsilon$ varies from 3.0 eV to 3.32 eV.
We demonstrate an absolute energy calibration model for the Ge detector response to low energy recoils. Two energy loss processes have been considered in the absolute calibration method compared to the traditional relative calibration, which only takes into account one energy loss process. Two energy loss processes a) and b) described above resulted in less ionization efficiency in the case of absolute calibration. Once all existing data were correctly scaled in the absolute energy response function, a good agreement between the theoretical prediction using the absolute energy response function and the data points was achieved. The difference between the absolute energy response function and the relative calibration energy scale in the visible energy is about a factor of $\sim$4, which will not result in a difference in the threshold of recoil energy within the calibration range because there are correspondences between the measured visible energy points and the incident recoil energy points. However, beyond the calibration range, the extrapolation of the energy scale using a relative calibration will posses large uncertainty up to a factor of 4 according to the absolute energy response function. 

\section*{Acknowledgments}
The authors would like to thank Christina Keller for her careful reading of this manuscript. We are also immensely grateful to Lauren Hsu and her collaborators in SuperCDMS for their comments on an earlier version of the manuscript and provide several references. In addition, the authors appreciate the wonderful suggestions and comments from Jing Liu and Chao Zhang. This work was supported in part by NSF PHY-0919278, NSF PHY-1242640, NSF OIA 1434142, DOE grant DE-FG02-10ER46709, and a governor's research center supported by the state of South Dakota.




\section*{References}
\begin{thebibliography}{99}
\bibitem{dm1} P. A. R. Ade et al. (Planck Collaboration), \emph{Planck 2013 results. I. Overview of products and scientific results}, \emph{Astronomy \& Astrophysics} {\bf 571} (2014) 48. arXiv:1303.5062v2.
\bibitem{dm2} M. W. Goodman and E. Witten, \emph{Detectability of certain dark-matter candidates}, \emph{Phys. Rev. D} {\bf 31} (1985) 3059.
\bibitem{dm3} R. J. Gaitskell, \emph{Direct Detection of Dark Matter}, \emph{Ann. Rev. Nucl. Part. Sci.} {\bf 54} (2004) 315.
\bibitem{dm4} B. W. Lee and S. Weinberg, \emph{Cosmological Lower Bound on Heavy-Neutrino Masses}, \emph{Phys. Rev. Lett.} {\bf 39} (1977) 165.
\bibitem{dan1} D. Barker and D.-M. Mei, \emph{Germanium Detector Response to Nuclear Recoils in Searching for Dark Matter}, \emph{Astropart. Phys.} {\bf 38} (2012) 1. arXiv:1203.4620v4.

\bibitem{mei} D.-M. Mei, W.-Z. Wei and L. Wang, \emph{Impact of Low-Energy Response to Nuclear Recoils in Dark Matter Detectors}, arXiv:1512.00694v1.
\bibitem{lind} J. Lindhard et al., \emph{Range Concepts
and heavy ion ranges (Notes on atomic collisions, II)}, \emph{Mat. Fys. Medd. Dan. Vid. Selsk.} {\bf 33} (14) (1963) 1.
\bibitem{dan2} D. Barker, W.-Z. Wei, D.-M. Mei and C. Zhang, \emph{Ionization efficiency study for low energy nuclear recoils in germanium}, \emph{Astropart. Phys.} {\bf 48} (2013) 8.
\bibitem{papp} T. Papp et al., \emph{A new approach to the determination of the Fano factor for semiconductor detectors}, \emph{X-Ray Spectrum} {\bf 34}(2005) 106-111.
\bibitem{birks1} J. B. Birks, \emph{The  efficiency of organic scintillators
}, \emph{Proc. Phys. Soc.} {\bf A64} (1951) 874. 
\bibitem{birks2} J. B. Birks, \emph{The Theory and Practice of Scintillation Counting} (1964). London: Pergamon.
\bibitem{mei2} D.-M. Mei et al., \emph{A model of nuclear recoil scintillation efficiency in noble liquids}, \emph{Astropart. Phys.} {\bf 30} (2008) 12.
\bibitem{lux} Jeremy J. Chapman (LUX Collaboration), Ph.D thesis, \emph{First WIMP Search Results from the LUX Dark Matter Experiment} (2014).

\bibitem{E&R} F. E. Emery and T. A. Rabson, \emph{Average energy expended per ionized electron-hole pair in silicon and germanium as a function of temperature}, \emph{Phys. Rev.} {\bf 140} (1965) A2089.
\bibitem{shock} W. Shockley, \emph{Problems related to p-n junctions in silicon}, \emph{Solid State Electronics} {\bf 2} (1961) 35.
\bibitem{pehl} R. H. Pehl et al., \emph{Accurate determination of the ionization energy in semiconductor detectors}, \emph{Nucl. Instrum. Meth.} {\bf 59} (1968) 45.
\bibitem{klein1} C. A. Klein, \emph{Semicondutor particle detectors: a research of the fano factor situation}, \emph{IEEE Transactions on Nuclear Science} {\bf 15} (1968) 214.
\bibitem{klein2} C. A. Klein, \emph{Bandgap dependence and related features of radiation ionization energies in semiconductors}, \emph{J. Appl. Phys.} {\bf 39} (1968) 2029. 
\bibitem{A&B} R. C. Alig and S. Bloom, \emph{Electron-Hole-Pair Creation Energies in Semiconductors}, \emph{Phys. Rev. Lett.} {\bf 35} (1975) 22.
\bibitem{superCDMS} R. Agnese et al. (SuperCDMS Collaboration), \emph{WIMP-Search Results from the Second CDMSlite Run}, \emph{Phys. Rev. Lett.} {\bf 116} (2016) 071301. arXiv:1509.02448v2.
\bibitem{fallows} S. M. Fallows (SuperCDMS Collaboration), Ph.D thesis, \emph{Measurement of Nuclear Recoils in the CDMS II Dark Matter Searc} (2014). 
\bibitem {fano} U. Fano, \emph{Ionization Yield of Radiations. II. The Fluctuations of the Number of Ions}, \emph{Phys. Rev.} {\bf 72} (1947) 26.


\bibitem{antman} S. O. Antman, D. A. Landis and R. H. Pehl, \emph{Measurements of the Fano factor and the energy per hole-electron pair in germanium }, \emph{Nucl. Instrum. Meth.} {\bf 40} (1966) 272.

\bibitem{smith} R. A. Smith, \emph{Semiconductors}, Cambridge University Press, London (1960).
\bibitem{varshni} Y. P. Varshni, \emph{Temperature dependence of the energy gap in semiconductors}, \emph{Physica} {\bf 34} (1967) 149.
\bibitem{thurmond} C. D. Thurmond, \emph{The standard thermodynamic functions for the formation of electrons and holes in Ge, Si, GaAs, and GaP}, \emph{Journal of the Electrochemical Society} {\bf 122} (1975) 1133.

\bibitem{cdmsprl} R. Agnese et al. (SuperCDMS Collaboration), \emph{Search for Low-Mass Weakly Interacting Massive Particles with SuperCDMS}, \emph{Phys. Rev. Lett.} {\bf 112} (2014) 241302.
\bibitem{edelweiss} E. Armengaud et al. (EDELWEISS Collaboration), \emph{Final results of the EDELWEISS-II WIMP search using a 4-kg array of cryogenic germanium detectors with interleaved electrodes}, \emph{Phys. Lett. B} {\bf 702} (2011) 329.
\bibitem{wang} G. Wang, \emph{Phonon emission in germanium and silicon by electrons and holes in
applied electric field at low temperature}, \emph{J. Appl. Phys.} {\bf 107} (2010) 094504.
\bibitem{neg&tro} B. S. Neganov and V. N. Trofimove, Otkrytia i izobreteniya {\bf 146} (1985).
\bibitem{luke} P. N. Luke, \emph{Voltage-assisted calorimetric ionization detector}, \emph{J. Appl. Phys.} {\bf 64} (1988) 6858.
\bibitem{ande} A. J. Anderson (SuperCDMS Collaboration), Ph.D thesis, \emph{A Search for Light Weakly-Interacting Massive Particles with SuperCDMS and Applications to Neutrino Physics} (2010).
\bibitem{HR} H. R. Zulliger and D. W. Aitken, \emph{IEEE Trans. Nucl. Sci.} {\bf 17} (1970) 187.
\bibitem{cdms_3} Z. Ahmed et al. (CDMS Collaboration), \emph{Analysis of the low-energy electron-recoil spectrum of the CDMS experiment}, \emph{Phys. Rev. D} {\bf 81} (2010) 042002.
\bibitem{lauren} Discussion with Lauren Hsu (SuperCDMS Collaborator) via email (Private communication).
\bibitem{edelweiss2} M. P. Chapellier et al., \emph{Physical interpretation of the Neganov-Luke and related effects}, \emph{Physica B} {\bf 284-288} (2000) 2135.
\bibitem{B&L} A. R. Beattie and P. T. Landsford, \emph{Auger effect in semiconductors}, \emph{ Proc. Roy. Soc.} (London) {\bf A249} (1959) 16.
\bibitem{cdms_2} Z. Ahmed et al. (CDMS Collaboration), \emph{Results from a Low-Energy Analysis of the CDMS II Germanium Data}, \emph{Phys. Rev. Lett.} {\bf 106} (2011) 131302.
\bibitem{nist} National Institute of Standards and Technology. Please refer to: {\it http://www.nist.gov/pml/data/star/}.
\bibitem{prin} Princetion Gamma Tech Inc.
\bibitem{pxi} X-ray Instruments Associates, 8450 Central Ave., Newark CA 94560, USA.
\bibitem{igor} Wavemetrics Inc., PO Box 2088, Lake Oswego, OR 97035, USA.
\bibitem{maj} R. J. Copper et al., \emph{Pulse shape discrimination for Gerda Phase I data}, \emph{Nucl. Instr. Meth. A} {\bf 629} (2011) 303.
\bibitem{gerda} M. Agostini et al., \emph{Pulse shape discrimination for Gerda Phase I data}, \emph{Eur. Phys. J. C} {\bf 73} (2013) 2583.
\bibitem{jones} K. W. Jones and H. W. Kraner, \emph{Stopping of 1- to 1.8-keV $^{73}$Ge Atoms in Germanium}, \emph{Phys. Rev. C} {\bf 4} (1971) 125.
\bibitem{collar} P. S. Barbeau, J. I. Collar and O. Tench, \emph{Large-Mass Ultra-Low Noise Germanium Detectors: Performance and Applications in Neutrino and Astroparticle Physics}, \emph{JCAP} {\bf 09} (2007) 009.
\bibitem{mess} Y. Messous et al., Astrophysics {\bf 3} (1995) 361-366.
\bibitem{chas} C. Chasman et al., \emph{Band-Gap Effects in the Stopping of Ge$^{72^{\star}}$ Atoms in Germanium}, \emph{Phys. Rev. Lett.} {\bf 21} (1968) 1430.
\bibitem{shutt} T. Shutt et al., \emph{Measurement of ionization and phonon production by nuclear recoils in a 60 g crystal of germanium at 25 mK}, \emph{Phys. Rev. Lett.} {\bf 69} (1992) 3425.
\bibitem{baud} L. Baudis et al., \emph{High-purity germanium detector ionization pulse shapes of nuclear recoils, gamma interactions and microphonism}, \emph{Nucl. Instrum. Methods Phys. Res. A} {\bf 418} (1998) 348.
\bibitem{simon} E. Simon et al., \emph{SICANE: a Detector Array for the Measurement of Nuclear Recoil Quenching Factors using a Monoenergetic Neutron Beam} \emph{Nucl. Instrum. Methods Phys. Res. A} {\bf 507} (2003) 643.
\end{thebibliography}


\end{document}